# The Influence of Spatial Discreteness on the Thermo-Diffusive Instability of Flame Propagation with Infinite Lewis Number

XiaoCheng Mi, Andrew J. Higgins, Samuel Goroshin, and Jeffrey M. Bergthorson
Department of Mechanical Engineering
McGill University, Montreal, Quebec, H3A 2K6 Canada

Corresponding author: Andrew J. Higgins
817 Sherbrooke Street West
Department of Mechanical Engineering
McGill University
Quebec, Canada H3A 2K6

e-mail: andrew.higgins@mcgill.ca

**Abstract**

The dynamics of flame propagation in systems with infinite Lewis number and spatially discretized sources of heat release is examined, which is applicable to the combustion of suspensions of fuel particles in air. The system is analyzed numerically using a one-dimensional heat equation with a source term for the reaction progress variable, which is specified to have zero diffusivity, and the model reveals a spectrum of flame-propagation regimes. For the case of a switch-type reaction rate and homogeneous media (continuous regime), the flame propagates steadily at a velocity in agreement with analytical solutions. As the sources are spatially concentrated into δ-function-like sources, propagation approaches the discrete regime with a fixed period between ignition of the sources, for which an analytic solution is also available for validation. When the source term is governed by an Arrhenius rate and the activation energy is increased beyond the stability boundary, the flame begins to exhibit a long-wavelength (4-5 times the thermal flame thickness) oscillation characteristic of the thermo-diffusive instability, in good agreement with prior stability analysis. When spatial discreteness is introduced, a competition is observed between the long-period oscillations of the thermo-diffusive instability and the pulsations associated with the rapid heat release of the concentrated sources. Interestingly, the presence of spatial discreteness is able to excite higher modes (period doubling and chaotic solutions) of the thermo-diffusive instability, suggesting that the introduction of discreteness may have an influence qualitatively similar to that of increasing activation energy. Relevance of the model parameters to experimental systems is then discussed.

# 1. Introduction

In many practical applications, incomplete mixing of fuel with oxidizer can result in the creation of inhomogeneous mixtures, wherein pockets of mixture that are able to burn alternate with regions that are effectively inert and flame propagation is not possible. Such systems may be created, for example, in the eddies of turbulent jets of fuel (gaseous or solid particulate) into an oxidizing atmosphere [1,2], incomplete blending of powdered reactants in pyrotechnic and Self-propagating High-temperature Synthesis (SHS) compositions, or produced intentionally, such as in pyrotechnic delay trains, in order to achieve very low burn rates [3]. Despite the presence of effectively inert gaps, many such systems can burn to completion via molecular transport through the inert regions, providing a delivery mechanism of the "ignition signal" between neighboring combustible pockets. In the case of a three dimensional system of randomly positioned pockets, an external observer may even observe what appears to be a continuous flame but may manifest some unusual propagation characteristics. To date, there have been a few attempts to study such systems, with the models that have appeared neglecting to consider the thermal gradients or flames inside the reactive pockets [4-6]. In some cases, this assumption is justified, such as in models of flame propagation in systems with highly spatially discrete sources. Such models are appropriate for treating systems of solid fuel particulates in suspension in gaseous oxidizer [4,6], since in these systems the particle size is negligible in comparison to the inter-particle spacing. The present study considers a more general case where the reactive pockets are not necessarily of negligible size.

In this paper, we consider spatially inhomogeneous, one-dimensional systems wherein reactive layers alternate in a regular array with layers that are nonreactive. For simplicity, we assume the diffusivity of the rate-limiting component is zero, as would be the case for a fuel-lean suspension of solid fuel in an oxidizing atmosphere where the particles remain in the condensed phase during the entire reaction and the global diffusion of the oxidizer is negligible, or in solid pyrotechnic or SHS compositions. Due to zero mass diffusivity, the distribution of fuel does not significantly change during the passage of the flame, which must rely solely upon thermal conduction in order to propagate. The regularly spaced suspension of nonvolatile fuel particles described by an ignition temperature "switch" was investigated in [4,5], and this scenario can be considered as the limiting, asymptotic case since the particles can be represented as spatial δ-functions. An analytic solution for flame propagation is available for this case [4], and thus can serve as verification of the numerical simulations presented in the current paper in the limit of highly discrete, reactive layers.

With the assumption of zero mass diffusivity of the rate limiting component, the Lewis number for the reactive system approaches infinity (Le → ∞), and as a result, flame propagation in a system that is homogeneous (i.e., continuum case) and described by an Arrhenius reaction rate is inherently unstable, as shown by Shkadinskii et al. [7]. Unlike mixtures with Le less than unity that exhibit a multidimensional cellular instability [8], the limit of large Le results in a one-dimensional oscillatory instability of the flame front for sufficiently large values of activation energy [9,10]. The flame front oscillations result in a spatial wavelength attributed to this inherent instability for the case of a homogeneous media. In the present study, the existence of alternating layers of reactive and inert media imposes an additional length scale to the problem, and the flame is expected to exhibit large pulsations as it traverses these intrinsic inhomogeneities in the system. In this study, the interplay between the oscillatory thermo-diffusive instability and the pulsating nature of flame propagation due to the alternating reactive layers is investigated as a function of the relative magnitude of these length scales, and as explored in detail below, displays a rich spectrum of different front dynamics.

## 2. Problem Description and Numerical Method

2.1 *Problem Description*

This problem is governed by the one-dimensional heat diffusion equation for temperature $T$ with a reaction source term and an equation tracking the reaction progress variable, $C$. The dimensionless form of the governing equations is as follows,

$$T_t = T_{xx} + \frac{R(T,C)}{\Gamma}, \quad C_t = -R(T,C) \tag{1}$$

For the switch-type model,

$$R(T,C) = \begin{cases} 0 & T < T_{ig} \text{ or } C = 0 \\ 1 & T \geq T_{ig} \text{ and } C > 0 \end{cases} \tag{2}$$

and for Arrhenius reaction rate,

$$R(T,C) = C(1-\sigma)\text{Exp}(-N/T) \tag{3}$$

The variables are non-dimensionalized as follows (note that tilde "~" indicates dimensional quantities). Time is $t = \tilde{t}/\tilde{t}_R$ where $\tilde{t}_R$ is the characteristic reaction time. Note that, for the Arrhenius reaction rate, $\tilde{t}_R$ is the inverse of the pre-exponential (or collision frequency) factor ($\sim 1/s$). The spatial coordinate is $x = \tilde{x}/\sqrt{\tilde{\alpha}\tilde{t}_R}$ where $\tilde{\alpha}$ is the thermal diffusivity. $C$ is the dimensionless reaction progress variable representing the concentration of reactant. The adiabatic flame temperature is $\tilde{T}_f = \tilde{T}_0 + \tilde{Q}/\tilde{\rho}\tilde{c}_p$ where $\tilde{T}_0$ is the initial temperature of the unburnt mixture and $\tilde{Q}$ is the average volumetric energy density of the system. For the switch-type model, temperature is $T = (\tilde{T} - \tilde{T}_0)/(\tilde{T}_f - \tilde{T}_0)$, and for an Arrhenius reaction rate, temperature $T = \tilde{T}/\tilde{T}_f$, initial temperature $\sigma = \tilde{T}_0/\tilde{T}_f$, and activation temperature $N = \tilde{E}_a/\tilde{R}\tilde{T}_f$, where $\tilde{E}_a$ is the activation energy, and $\tilde{R}$ is the universal gas constant. Note that no mass diffusion appears in the governing equations, i.e., Le = ∞ in the proposed system. An analytic solution for flame speed $\widetilde{V}_f$ with switch-type and Arrhenius reaction rate can be found in [4,6] and [11], respectively. Knowing $\widetilde{V}_f$, the thermal thickness of flame can be estimated as $\widetilde{l}_r = \widetilde{\alpha}/\widetilde{V}_f$, and in dimensionless form, $l_r = 1/V_f$.

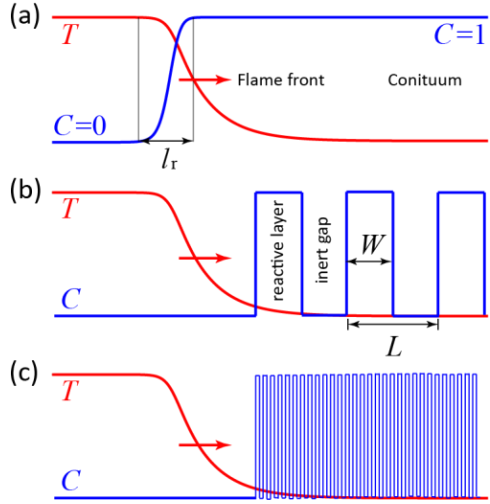

Figure 1: Schematic representation of the problem.

The reactive medium is spatially discretized in a one-dimensional system by concentrating reactant into layers separated by inert gaps. As shown in Fig. 1, the discretization can be realized by simply initializing the reaction progress variable $C$ as 1 (i.e., unreacted) in the reactive layers and 0 (i.e., no reactant) in the inert gaps. The discreteness of the system is described by a parameter $\Gamma = W/L$, where $W$ is the reactive layer width and $L$ the spacing between two consecutive layers. The volumetric energy density in each discrete layer is $\widetilde{Q}/\Gamma$, so that the average volumetric energy density of the overall medium can be maintained at $\widetilde{Q}$ while $\Gamma$ is varied. Discreteness on a scale comparable to (Fig. 1b), or much smaller than (Fig. 1c), the flame thickness can be introduced by independently varying $\Gamma$ and $L$. Volumetric dilation (i.e., decrease in density) due to the heat release is approximately compensated for by an increase in thermal conductivity of the medium at higher temperature. Hence, for simplicity, we propose to neglect these competing effects in this problem for the initial study reported here.

For switch-type reaction rate, the simulations were performed with an ignition temperature $T_{ig} = 0.4$ that is representative of some heterogeneous combustion systems (e.g., lean suspensions of iron powders with $\widetilde{T_{ig}} = 900$ K and $\widetilde{T_f} = 2200$ K [12]). For an Arrhenius reaction rate, $\sigma$ and $N$ are the ratios of the initial temperature and activation temperature, respectively, to the adiabatic flame temperature. The calculations reported in this paper were performed with $\sigma = 0.1$ and $N = 6$ and $8$, which are representative of the combustion of micro-sized metal particles (e.g., $\widetilde{T_0} = 300$ K, $\widetilde{T_f} = 3451$ K and 2857 K, and $\widetilde{E_a}/\widetilde{R} = 16000$ K and 22600 K, for aluminum and boron particles, respectively [13,14]).

2.2 *Numerical Method*

A numerical scheme of central differencing in space and explicitly forward differencing in time was used to solve Eqs. (1)-(3). Temperature was uniformly initialized as $T = 0$ for the switch-type model and $T = \sigma$ for an Arrhenius reaction rate while $C$ was initialized as piecewise step functions according to the prescribed $\Gamma$ and $L$. A hot wall maintained at the adiabatic flame temperature $T = 1$ at the left boundary was used to initiate a flame propagating rightward. In order to avoid any influence on the flame dynamics imposed by the boundary condition, the flames were simulated to propagate over a distance nearly 100 times the average thermal thickness of the flame in the continuous case ($l_r$). At least 100 computational cells per $l_r$ and 6 cells per reactive layer ($W$) were used in this study. The position of the flame front $x_f(t)$ was recorded at each time step by finding the location where $T$ first increases to $T_{ig}$ for switch-type model or $(1 + \sigma)/2$ for Arrhenius reaction rate, and the instantaneous flame velocity $V_f$ was obtained by numerically differentiating $x_f(t)$ using a central differencing scheme.

## 3. Results and Analysis

3.1 *Switch-type reaction rate*

With $T_{ig} = 0.4$, flame propagation is stable in both the continuous and discrete systems, so analytic solutions for the flame speed can be found in the case of a homogeneous media and in the limit of highly discrete, δ-function-like sources (i.e., $\Gamma \to 0$) [4,6].

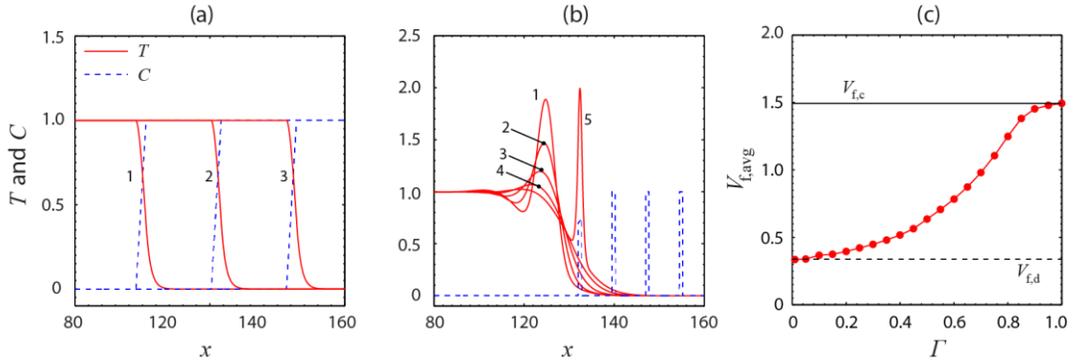

Figure 2: Sample results of switch-type reaction-rate model, with $T_{ig} = 0.4$, for the (a) continuous ($\Gamma = 1$) and (b) discrete ($L = 5l_r$ and $\Gamma = 0.1$) cases. (c) Simulation results of average flame speed as a function of spatial discreteness compared to analytic solutions for continuum (solid line) and in the limit of $\Gamma \to 0$ (dashed line). The numeric labels in (a) and (b) indicate the chronological order of the snapshots of $T$ and $C$ profiles.

In Fig. 2a, selected snapshots of temperature and reaction progress variable profiles with a steady shape show a stable flame propagating in a continuous system ($\Gamma = 1$). The sample results of $T$ and $C$ profiles (labeled from 1 to 5) of a flame propagating in a highly discrete system are shown in Fig. 2b. In Fig. 2c, the average flame speed $V_{f,avg}$ measured from the simulations is plotted as a function of discreteness $\Gamma$ while layer spacing $L$ is held at a constant value of $5l_r$. After the flame has propagated through the first 5 reactive layers, there is no significant change in the average speed at which the flame front travels from one layer to the next. The simulation results of $V_{f,avg}$ plotted in Fig. 2c were calculated from the 15$^{th}$ to the 20$^{th}$ reactive layer. As shown in Fig. 2c, $V_{f,avg}$ gradually increases as the discreteness parameter $\Gamma$ increases from 0.01 to 1, reaching plateau values at both $\Gamma \to 0$ and $\Gamma = 1$ extrema. These two plateaus tend to approach the analytic solutions of the flame speed for the continuous case $V_{f,c} = 1.4983$ and for the highly discrete case $V_{f,d} = 0.3372$, both of which are plotted as horizontal lines in Fig. 2c. The agreement between simulation results and theoretical predictions validate the numerical method and the technique of introducing spatial discreteness used in this study.

## 3.2 Arrhenius reaction rate

### 3.2.1 Sample results

The results of sample calculations with an Arrhenius reaction rate are shown in Fig. 3. Selected snapshots (labeled from 1 to 5) of both temperature and reaction progress variable profiles showing a flame front propagating rightward are plotted in Fig. 3. Three qualitatively distinct scenarios can be identified here: Fig. 3a shows a stable flame propagating in a continuous system ($N = 6$, $\sigma = 0.1$, and $\Gamma = 1$); Fig. 3b is a weakly oscillating flame propagating in a continuous system ($N = 8$, $\sigma = 0.1$, and $\Gamma = 1$); Fig. 3c is a severely pulsating flame propagating in a highly discrete medium ($N = 8$, $\sigma = 0.1$, $\Gamma = 0.1$, and $L = 500$).

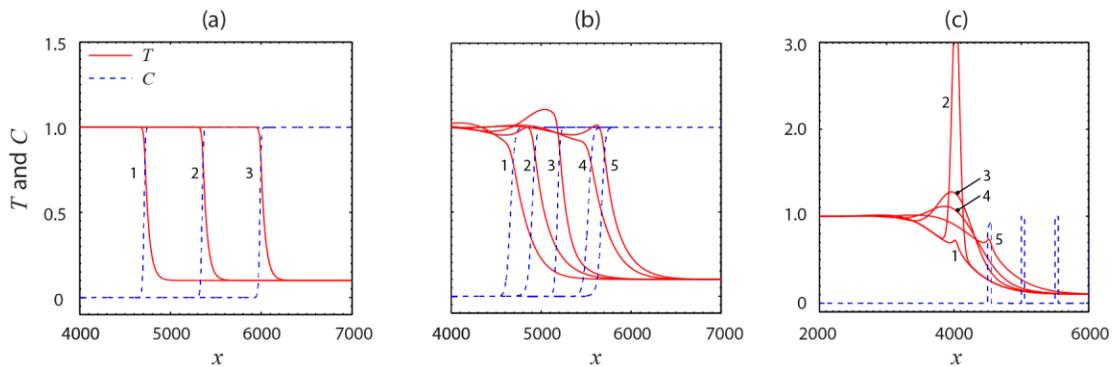

Figure 3: Sample results of Arrhenius reaction-rate model for (a) continuous, stable ($\sigma = 0.1$, $N = 6$, and $\Gamma = 1$), and (b) unstable ($\sigma = 0.1$, $N = 8$, and $\Gamma = 1$), and (c) discrete ($\sigma = 0.1$, $N = 8$, $L = 500$, and $\Gamma = 0.1$) cases. The numeric labels indicate the chronological order of the snapshots of $T$ and $C$ profiles.

In Fig. 3a, the shape of the $T$ and $C$ profiles remains unchanged over time, and the flame front advances nearly the same distance after each constant time interval. Thus, a steady flame speed is calculated from the simulation result as $V_f = 0.0213$, which agrees remarkably well with estimated value $0.0214$ using the analytic solution in [11]. This result provides a control case which exhibits no thermo-diffusive instability with a sufficiently low activation energy of $N = 6$.

In Fig. 3b, by increasing the activation energy to $N = 8$ in comparison to Fig. 3a, the resulting temperature profile begins to exhibit an oscillatory behavior. In each profile, small "humps" and "valleys" can be observed as the temperature near the reaction zone fluctuates $\pm 10\%$ around the adiabatic flame temperature. Although no steady flame speed can be obtained in this case, an average flame speed can be calculated as $V_{f,avg} = 0.0063$, which is still close to the theoretical estimation $V_{f,c} = 0.0068$ [11]. An average thermal thickness of the flame can be calculated as $l_r = 1/V_{f,avg} \approx 160$. Similar oscillating flame propagation has been observed and attributed to the thermo-diffusive instability of the large Le limit in [7,10]. The results shown in Fig. 3a and b suggest that the stability boundary for $\sigma = 0.1$ is between $N = 6$ and 8, which is fairly close to the theoretical prediction of $N \approx 9.4$ in [9].

In Fig. 3c, only a single profile of reaction progress variable is plotted. At the instant corresponding to this snapshot (profile 5), a layer of reactant is being consumed while those ahead of it remain nearly unreacted. The temperature profiles exhibit a more severely pulsating behavior, with peak temperature occasionally exceeding three times of the adiabatic flame temperature. This more intensely pulsating propagation is only excited as discreteness is introduced to the system. The local peak temperature (in profile 2) is due to the rapid energy release of the highly concentrated reactant in a discrete layer. Similar results in discrete systems have been discovered via numerical simulations performed by Krishenik et al. [15].

3.2.2 Regular flame oscillations

As a set of parameters (i.e., $\sigma = 0.1$ and $N = 8$ in the rest of this paper) which give rise to thermo-diffusive instabilities in a continuous system (as shown in Figs. 3b and 3c) are imposed to a discrete medium, the flame propagation might be influenced by both the intrinsic thermo-diffusive instability and the spatial discreteness. In order to distinguish the regime of instability of the resulting flame pulsation, it is of interest to examine the power spectral density (PSD) obtained by performing a Fast Fourier Transform (FFT) of the history of the instantaneous flame velocity as a function of flame position (i.e., $V_f(x)$). Note that, since the objective of this analysis is to examine how flame pulsation is influenced by the spatial discreteness, it is more convenient to consider the flame pulsation as a spatial signal and characterize the oscillatory behavior in terms of wavelength rather than temporal frequency. Subfigures in the upper row in Fig. 4 are the plots of $V_f(x)$ for cases with various values of $\varGamma$ and $L$, while those in the lower row are the corresponding plots of PSD over the wavelength $\lambda$ domain.

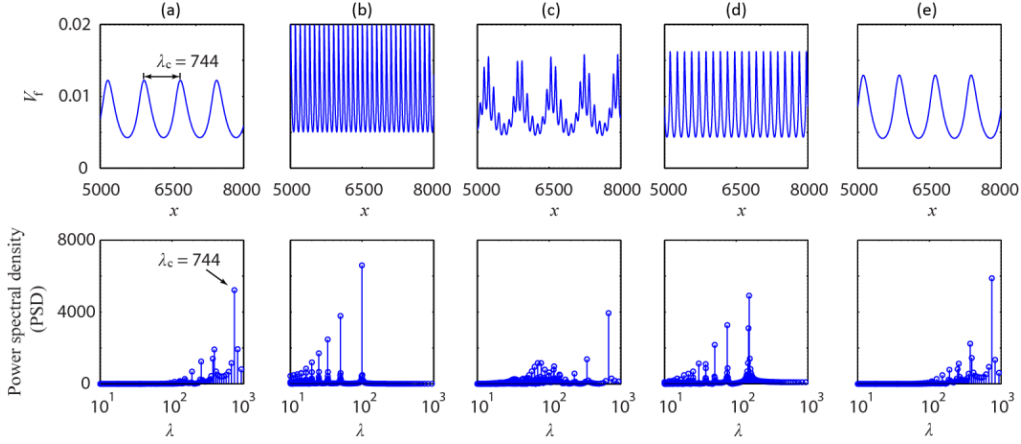

Figure 4: Simulation results of Arrhenius reaction-rate model ($\sigma = 0.1$ and $N = 8$). Upper row: history of the instantaneous flame velocity as a function of position. Lower row: PSD obtained via FFT plotted in wavelength domain for the following cases: (a) continuous, (b) $L = 100$, $\Gamma = 0.1$, (c) $L = 100$, $\Gamma = 0.5$, (d) $L = 150$, $\Gamma = 0.5$, and (e) $L = 30$, $\Gamma = 0.5$.

The continuous case of $\sigma = 0.1$ and $N = 8$ is shown in Fig. 4a. The results of $V_\text{f}(x)$ exhibit a regular oscillation with one single period. Thus, a unique wavelength of this oscillation, $\lambda_\text{c} = 744$, which is about 4.65 times the average $l_\text{r}$, can be easily identified from the plot of $V_\text{f}(x)$. Since no steady flame structure exists for this unstable system, it is more convenient to use this $\lambda_\text{c}$ as the characteristic length scale of the intrinsic thermo-diffusive instability, rather than an estimated or averaged $l_\text{r}$. Note that, in the PSD plot, the smaller-amplitude peaks at half, one fourth, and so on, of $\lambda_\text{c}$ are the harmonics of the dominant oscillation mode, while the non-zero PSD between these harmonic wavelengths is numerical noise.

The discrete case of $L = 100$ ($\approx 0.13\lambda_\text{c}$) and $\Gamma = 0.1$ shown in Fig. 4b also exhibits a regular single-period oscillation with an amplitude much larger than that of the continuous case and a much shorter wavelength. From the PSD plot, the dominant wavelength of this oscillation is 100, which is the same as the prescribed layer spacing $L$. In Fig. 4c, increasing $\Gamma$ to 0.5 while holding $L = 100$, the signal $V_f(x)$ exhibits a small-amplitude, short-wavelength oscillation ("small ripples") on top of a large-amplitude, long-wavelength oscillation ("large humps"). A spike in the PSD at $\lambda = 692$ ($\approx 0.93\lambda_c$) is clearly the wavelength associated with these large humps. A significantly non-zero PSD is observed in a band near a wavelength $\lambda = 100$, which is equal to $L$ in this case. Although more than one dominant wavelength is found in such a flame pulsation, it is still considered as regular since there is only one dominant wavelength on the scale of $\lambda_c$, which is clearly distinguishable from those on the scale of $L$.

Holding the discreteness at $\Gamma = 0.5$ and increasing $L$ to $150$ ($\approx 0.2\lambda_c$), as shown in Fig. 4d, the signal of $V_f(x)$ again exhibits an oscillation with short wavelength. From the PSD plot, a single dominant wavelength (with its harmonics) can be found at $\lambda = 150$, which is the same as $L$ in this case. Still holding $\Gamma = 0.5$ but decreasing $L$ to $30$ ($\approx 0.04\lambda_c$), as shown in Fig. 4e, the signal of $V_f(x)$ reverts to a single-period, smooth oscillation with a long wavelength. The PSD plot shows that this dominant wavelength is nearly the same as $\lambda_c$. The PSD in the wavelength band near the layer spacing ($L, \lambda = 30$) is nearly zero.

3.2.3 Flame pulsations with period doubling and highly irregular behavior

For some highly discrete cases with small $L$, the resulting flame pulsation is no longer a regular oscillation, but exhibits either period-doubling or chaotic behavior. In the case of $L = 50$, as $\Gamma$ decreases to 0.05, two distinguishable oscillation modes with wavelengths on the scale of $\lambda_c$ begin to appear. In Fig. 5a, two different peak values associated with the large humps are indicated by the horizontal dashed lines. In this case, the short-wavelength oscillations on the scale of $L$ can also be observed on top of the period-two, long-wavelength oscillation.

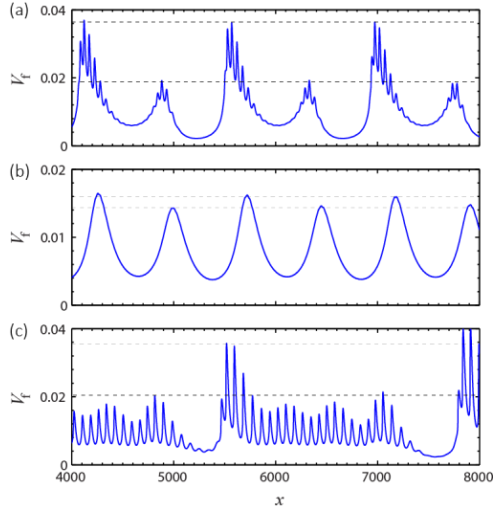

Figure 5: Simulation results of the history of the instantaneous flame velocity as a function of position with Arrhenius reaction-rate model ($\sigma = 0.1$ and $N = 8$) for the cases of $\Gamma = 0.05$ and $L =$ (a) 50, (b) 30, and (c) 80.

While holding $\Gamma = 0.05$ and decreasing $L$ to 30, the history of $V_f(x)$ also exhibits a period-two oscillation with wavelengths on the scale of $\lambda_c$, and two distinguishable peak values of $V_f$ can be clearly seen in Fig. 5b as indicated by the horizontal dashed lines. In this case with smaller $L$, the oscillation with a wavelength on the scale of $L$ can no longer be observed in the signal of $V_f(x)$ plotted in Fig. 5b. Holding $\Gamma = 0.05$ and increasing $L$ to 80, $V_f(x)$ plotted in Fig. 5c shows that, although the short-wavelength (on the scale of $L$) oscillation appears to be relatively regular in some regions, the oscillatory behavior on the long-wavelength scale is highly irregular. From the plot, two peak values of $V_f$ can be found and indicated by the dashed lines. Whether these peaks are associated with a period-two limit cycle or the flame pulsation eventually evolves to a chaotic solution remains unknown until results with a much longer domain are obtained.

## 4. Discussion

The discussion is first focused on the results of the moderately discrete cases (i.e., $0.5 \leq \Gamma < 1$). Among these cases, three clearly distinct regimes of flame pulsation can be identified: effectively-continuous, discrete, and mixed regimes. Certainly, as $\Gamma$ approaches a value of unity, the system reverts to a continuum wherein the long-wavelength ($\lambda_c$) flame oscillation reflects the thermo-diffusive instability of an infinite Lewis number system. For a moderately-discrete system, as $L$ decreases to about 5% of $\lambda_c$ (the case shown in Fig, 4e), the resulting flame dynamics revert back to the behavior of the continuous case. The system is thus "homogenized" by such fine scale heterogeneities and gives rise to flame propagation under the so-called effectively-continuous regime. A similar effect is encountered in the problem of detonation propagating in a heterogeneous medium, where it was found that, as the size of the heterogeneity was made less than the half-reaction-zone length, the resulting detonation speed and structure reverts to that of the homogeneous case [16].

As $L$ increases to about 20% of $\lambda_c$, but still in the moderately-discrete regime, the flame pulsation appears entirely dominated by the frequency of discrete layers releasing heat (e.g., Fig. 4d). In such a case, the thermo-diffusive instabilities that might be triggered inside the reactive layer are completely damped out as heat diffuses through the long, inert gaps between them. Thus, only the rapid burning of the reactant concentrated in the discrete layer can exert a strong pulse on the propagating flame. Such flame propagation behavior is fully within the discrete regime.

Between the discrete and effectively-continuous regimes, flame propagation can be significantly influenced by the pulses of the discrete layers releasing their heat while the intrinsic thermo-diffusive instabilities are also present. Such a case is considered to be of a mix of thermo-diffusive and discrete pulsations and exhibits oscillatory behaviors of two main frequencies, as shown in Fig. 4c.

For a moderate layer spacing (i.e., $100 \leq L \leq 200$), as $\Gamma$ decreases from 1 to the limit of 0, the resulting flame propagation transits through the three above-mentioned regimes: continuous, mixed, and discrete regimes. The simulation results have been categorized according to their propagation regimes and summarized as a map in $L$-$\Gamma$ parameter space (Fig. 6). Note that, in Fig. 6, the boundaries of the colored regions designating various propagation regimes are not quantitative simulation results, but rather a qualitative, visual guide to assist the reader in identifying different categories.

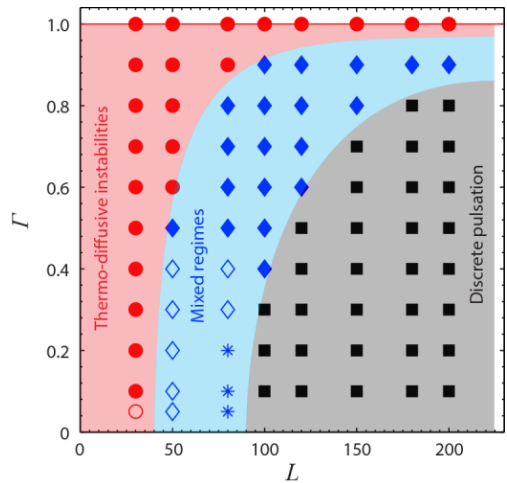

Figure 6: Simulation results categorized according to the flame propagation regimes and summarized as a map in $L$-$\Gamma$ parameter space where solid circle represents effectively-continuous regime, solid square discrete regime, solid diamond mixed regime, asterisk likely chaotic, and the open symbols indicate period doubling.

A more complicated flame propagation behavior was found in the region of small $L$ and small $\Gamma$ (i.e., bottom-left corner of Fig. 6). Examining at the column of $L = 50$ in Fig. 6, $\Gamma = 0.4$ marks the onset of the period-doubled mode, while the presence of the mixed regime is still identifiable in this case. Interestingly, by making the system more spatially discrete, an extra mode of oscillation attributed to the intrinsic thermo-diffusive instability of the continuous system can be excited. For $L = 30$, although the effect of discreteness on the flame dynamics is not visible as short-wavelength oscillations for any value of discreteness, a second mode of oscillation with a wavelength on the scale of $\lambda_c$ is triggered as $\Gamma$ decreases to $0.05$. Also as a consequence of decreasing $\Gamma$, for $L = 80$, a highly irregular flame oscillation is triggered, which exhibits the characteristics of being fully chaotic. Although the complex nonlinear dynamics of these irregularly pulsating flames cannot be extensively studied within the scope of this paper, these preliminary results suggest that spatial discreteness may have a destabilizing effect, similar to increasing activation energy, on flame dynamics.

## 5. Conclusion

Unlike the previous theoretical and numerical investigations of discrete-source combustion [4-6,15], spatially finite reactive elements have been introduced to a thermo-diffusively unstable combustible medium in this paper. The results obtained via numerical simulation have demonstrated that the introduction of spatial discreteness of the heat sources has a significant influence on the resulting flame instability. The intrinsic thermo-diffusive instabilities can be damped or excited by the introduction of discrete heat sources, depending on their spatial discreteness and spacing in comparison to the wavelength of these intrinsic oscillations. To consider an experimental system, the observed flame pulsations in lean suspensions of nonvolatile aluminum particulates in oxidizing atmosphere occur at a wavelength of approximately 2-5 mm, while the reported average particle spacing is 0.12 mm [12], leading to a value of $L \approx$ 20-60 mm in the model proposed here. As a result, the flame propagation is very likely influenced by a regime of thermo-diffusive instability and partially mixed with discrete pulsation. For a lean suspension of aluminum particles in a balloon [2] with a similar average particle spacing (0.25 mm) to that studied in [12], the observed spherically pulsating flame might also be governed by this mixed regime identified in this paper. This idealized, one-dimensional model outlines an approach to systematically studying flame pulsation in a discrete reactive medium influenced by interacting physics over multiple length scales. In order to fully capture the effect of discrete combustion sources, their random distributions in three-dimensional space must be considered in future efforts.